\documentclass{IEEEtran}
\usepackage{cite}
\usepackage{amsmath,amssymb,amsfonts}
\usepackage{algorithmic}
\usepackage{graphicx}
\usepackage{textcomp}
\def\BibTeX{{\rm B\kern-.05em{\sc i\kern-.025em b}\kern-.08em
    T\kern-.1667em\lower.7ex\hbox{E}\kern-.125emX}}
\makeatother
\usepackage{color}
\definecolor{mycolor}{rgb}{1,0.6470588235294117,0.6980392156862745}
    
\begin{document} 
\twocolumn


\title{
Protected Transverse Electric Waves in Topological Dielectric Waveguides}
\author{Rui Zhou, Menglin L. N. Chen, Xintong Shi, Yan Ren, Zihao Yu, Yu Tian, Y. Liu, Hai Lin
\thanks{ R. Zhou, X. T. Shi, Y. Ren, Z. H. Yu, and H. Lin are with College of Physics Science and Technology, Central China Normal University, Wuhan 430079, Hubei Province (e-mail: linhai@mail.ccnu.edu.cn). }
\thanks{M. L. N. Chen and R. Zhou are with the Department of Electrical and Electronic Engineering, The Hong Kong Polytechnic University, Hong Kong (e-mail: menglin.chen@polyu.edu.hk)}
\thanks{ Y. Tian and Y. Liu are with Department of Physics, School of Physics, Hubei University, Wuhan 430062, Hubei Province, and  Y. Liu also with Lanzhou Center for Theoretical Physics, Key Laboratory of Theoretical Physics of Gansu Province, and Key Laboratory of Quantum Theory and Applications of MoE, Lanzhou University, Lanzhou, Gansu 730000}
 }

\maketitle
\begin{abstract}
Waveguides are fundamental components in communication systems. However, they suffer from reflection and scattering losses at sharp routes or defects. The breakthrough in developing topological photonic crystals (PhCs) provides promising solutions to robust signal transmission. In this work, we propose a new mechanism for protecting wave-guiding modes by decorating the boundaries of a conventional waveguide with valley-Hall PhCs. This special layout enables the robust propagation of conventional transverse electric waves against defects and bends. Moreover, the proposed waveguide is compatible with the substrate integrated waveguide (SIW). High efficient mode conversion from the SIW to the proposed waveguide is achievable. By leveraging the idea of topology to conventional waveguides, we provide a powerful and practical tool that can largely improve the performance of microwave and millimeter-wave integrated circuits while reserving the features of wave-guiding modes.
\end{abstract}

\begin{IEEEkeywords}
Valley photonic crystals, millimeter-wave integrated waveguides, robust wave propagation, topological waveguides.
\end{IEEEkeywords}

\section{Introduction}
\label{sec:introduction}
Wave-guiding structures are carriers for the transportation, modulation, and processing of electromagnetic (EM) signals. Since the next-generation communication system exploits higher frequencies, waveguides operating at millimetre-wave (mmWave) regimes have been extensively explored and applied due to their low loss and high power handling capabilities~\cite{2003SingleDeslandes,2005GuidedXu,2015DesignBrazalaz,2021ShenmmW,2022GaOn,2019BautistaCompact}. Because of the commonly existing discontinuities in actual circuits, the degradation of signals will inevitably appear at defects or intersections when there is a mode mismatching~\cite{2014BroadbandMa,2019AdiabaticXu}. Research has been conducted to suppress the mode mismatching in bended waveguides~\cite{2017StudyZhang}, but the designs are not adaptable. 

The discovery of topological photonic crystals (PhCs) offers a promising outlook for robust signal transmission. The topological phase of matter which was first discovered in electronic systems has been extended to electromagnetics~\cite{2013PhotonicKhanikaev,2008PossibleHaldane,2008ReflectionWang,2015SchemeWu,2016ObservationLu,2022HigherZhou,2022lowLi,2011ExperimentalPoo}. The most striking feature arising from the topological phase is that the induced topological edge states can propagate robustly against local system deformations. In topological photonics, the spin-Hall and valley-Hall PhCs have been developed to provide innovative platforms to support topologically protected EM transmission~\cite{2017ValleyleChen,2018TunableChen,2018YangVisualization,2021TopologicalZhou,2019PseudospinChen, 2022TopologicalYu}. The proposed topological waveguides are prominent in integrated circuits for on-chip communication~\cite{2020TerahertzYang,2019TopologicalMa,2017ValleyDong,2018ValleyChenQ,2011DirectWu,2018ValleyZhang,2022ValleyLi}.  
However, topological edge states are usually tightly confined at domain walls, which makes them not compatible with conventional waveguiding modes. Although the recently reported three-layer heterostructures can introduce an additional width degree of freedom into topological waveguides by adding a middle domain featuring a Dirac cone, the topological modes being supported are still chiral states~\cite{2020ValleyWang,2022ExtendedWang,2021TopologicalWang,2021PhotonicChen,2023ExtendedZhao,2023Coexistenceli}. They do not share any similarity with the modes in conventional waveguides. The efficient excitation of these topological devices relies on chiral source arrays, hindering their integration into existing photonic circuits.

To solve the above mentioned problems, we propose a novel topological valley-locked waveguide (TVLW) that consists of an adjustable air channel in the middle of two valley-Hall PhCs with opposite topological phases. The proposed TVLW supports two distinct modes simultaneously: one is the conventional transverse electric (TE) mode supported at the core-air channel, and the other is the valley-locked edge states appearing on both sides of the air channel. Since the proposed TVLW has an additional width degree of freedom, it can be interconnected with conventional substrate-integrated waveguides (SIWs) by adjusting the channel width,  which has never been reported before. Based on the transmission simulation, we successfully prove that the hybrid mode presents the topological protection feature when there are defects and bends. Our design is the first effort to integrate topological robust transmission to conventional wave-guiding modes, opening the door to the practical implementation of topology features in integrated circuits.

\section{\label{sec:level1}Theory and model}

\subsection{Photonic crystal design}

A PhC arranged in a triangular lattice is shown in Fig.~1(a). The unit cell consists of dielectric rods (dielectric constant 11.2) with the height of $h_2=1.08$ mm. For ease of fabrication, a slab ($h_1-h_2=0.2$ mm, dielectric constant $11.2$) is reserved under the dielectric rods. The whole structure is covered by the top and bottom metal plates, as shown in the upper panel of Fig.~1(a). The black rhombus or hexagon in the bottom panel shows the unit cell which consists of two types of rods, A and B. Their diameters are denoted by $d_{\mathrm{A}}$ and $d_{\mathrm{B}}$, respectively. The lattice vectors are represented by $a_{\mathrm{1}}$ and $a_{\mathrm{2}}$ with the lattice constant $a_0=5$~mm. The finite element method is used to simulate the PhC. As shown in Fig. 1(b), when $\Delta d=0\left(\Delta d=d_{\mathrm{A}}-d_{\mathrm{B}}, d_{\mathrm{A}}=d_{\mathrm{B}}=0.48 a_0\right)$, band degeneracy occurs at the $K / K^{\prime}$ points at two frequencies (black dashed lines). When the inversion symmetry of the unit cell is broken by setting $\Delta d \neq 0$, two complete band gaps will appear (black solid lines). By setting $\Delta d=\pm 0.16 a_0 (d_{\mathrm{A}}=0.56 a_0, d_{\mathrm{B}}=0.40 a_0, \text { or } d_{\mathrm{A}}=0.40 a_0, d_{\mathrm{B}}=0.56 a_0$), there are two band gaps at $14.1-15.4$~GHz ($f_{\mathrm{1}}$) and $28-32$~GHz ($f_{\mathrm{2}}$).

\begin{figure}[!htbp]
\centering\includegraphics[width=3.4in]{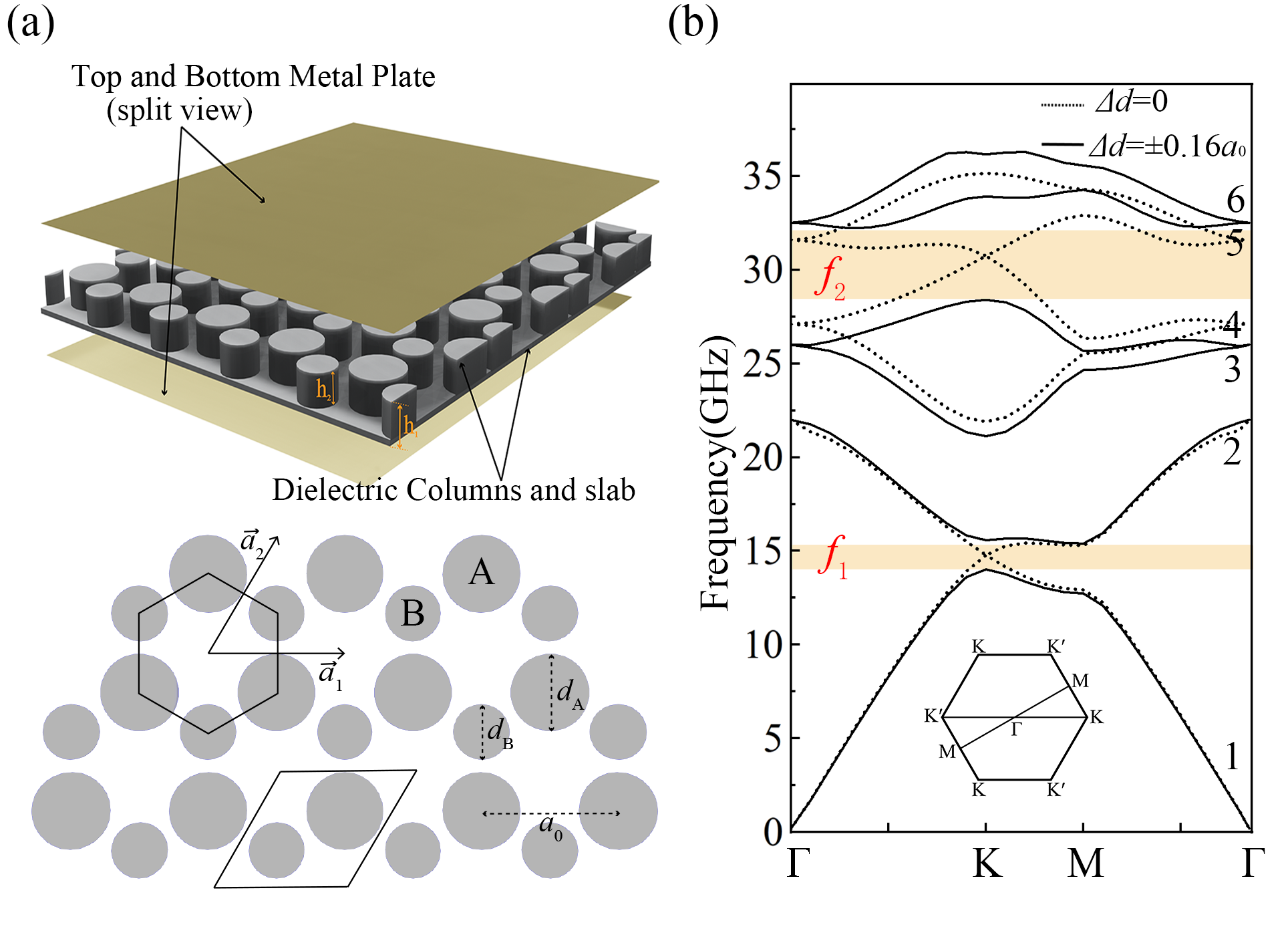}
\caption{A PhC with subwavelength thickness and its bulk band. (a) Upper panel: Three dimensional split view of the PhC (Note that there is no gap between the three layers of the structure). The top and bottom are metal plates. $h_1=1.28$ mm, $h_2=1.08$ mm. Bottom panel: The PhC lattice. Both the black rhombus and hexagon represent unit cells. The unit cell vectors are represented by $a_{\mathrm{1}}$ and $a_{\mathrm{2}}$. The lattice constant is $a_0=5$ mm, $d_{\mathrm{A}}$ and $d_{\mathrm{B}}$ represent the diameters of two different rods. (b) The dispersion diagram of PhCs when $\Delta d=0$ (dashed lines) and $\Delta d=\pm 0.16 a_0$ (solid lines), where the orange areas, $f_{\mathrm{1}}$ and $f_{\mathrm{2}}$ represent the two band gaps.}
\end{figure}

The properties of the states near the $K$ valley at the four bands that make up the two band gaps are explored. When $\Delta d=0.16a_0$, the phase of $E_z$ and corresponding Poynting vectors are shown in Fig.~2(a). For band 1 and band 5, the $E_z$-phase experiences a $2\pi$ decrease anticlockwise around the center of the B-rod, and the Poynting vector shows that it has a left-handed circularly polarized (LCP) phase vortex. For band 2 and band 4, the $E_z$-phase decreases $2\pi$ clockwise around the center of the A-rod, and the Poynting vector shows that it has a right-handed circularly polarized (RCP) phase vortex. This vortex phase distribution is a typical feature of valley-Hall PhCs~\cite{2019QuantizationBenalcazar}. 

Then, we use the Berry curvature to classify the topological states of the PhCs. The Berry connection of the four bands (bands 1, 2, 4, 5) is defined by~\cite{2017ValleyleChen,2011DirectOkada}:

\begin{equation}
\vec{A}(\vec{k}) \equiv i\left\langle u_{\vec{k}}\left|\nabla_{\vec{k}}\right| u_{\vec{k}}\right\rangle=i \oint_{{unitcell}} {\rm d}^{2} r \varepsilon(\vec{r}) u_{\vec{k}}^{*}\left[\nabla_{\vec{k}} u_{\vec{k}}\right]
\end{equation}
where $u_{\vec{k}}$ is the normalized eigenstate $\left\langle u_{\vec{k}} \mid u_{\vec{k}}\right\rangle=1$ and the asterisk denotes complex conjugation. $\varepsilon(\vec{r})$ is the spatial permittivity distribution. The gauge independent Berry curvature, $\Omega(\vec{k})$ is obtained by:
\begin{equation}
\Omega(\vec{k}) \equiv \nabla_{\vec{k}} \times \vec{A}(\vec{k})=\frac{\partial A_{y}(\vec{k})}{\partial k_{x}}-\frac{\partial A_{x}(\vec{k})}{\partial k_{y}}. 
\end{equation} Figure 2(b) shows the calculated Berry curvature near the $K$ valley of the four bands, which is opposite to that near the $K^{\prime}$ valley (not shown here). Topological indices $C_{\mathrm{K} / \mathrm{K}^{\prime}}$ at the $K$ and $K^{\prime}$ valleys are defined as the integration of Berry curvature within a half First Brillouin Zone, and the valley-Chern number is defined as $C_{\mathrm{V}}=\left|C_{\mathrm{K}}-C_{\mathrm{K}^{\prime}}\right|$\cite{2020FirstZhao}. As a result, the valley-Chern number $C_{\mathrm{V}}=|\frac{1}{2}-\left(-\frac{1}{2}\right)|=1$ for band 1 and 5, and  $C_{\mathrm{V}}=-1$ for band 2 and 4. By changing $\Delta d$, it is observed that the chirality of the phase for adjacent frequency bands is reversed at the critical point $\Delta d=0$, and blue (red) arrows indicate LCP (RCP) chiral states. The chirality reversal of the phase indicates the topological phase transition, which is proved by the opposite valley-Chern numbers in Fig.~2(c).

\begin{figure}[!t]
\centering\includegraphics[width=3.3in]{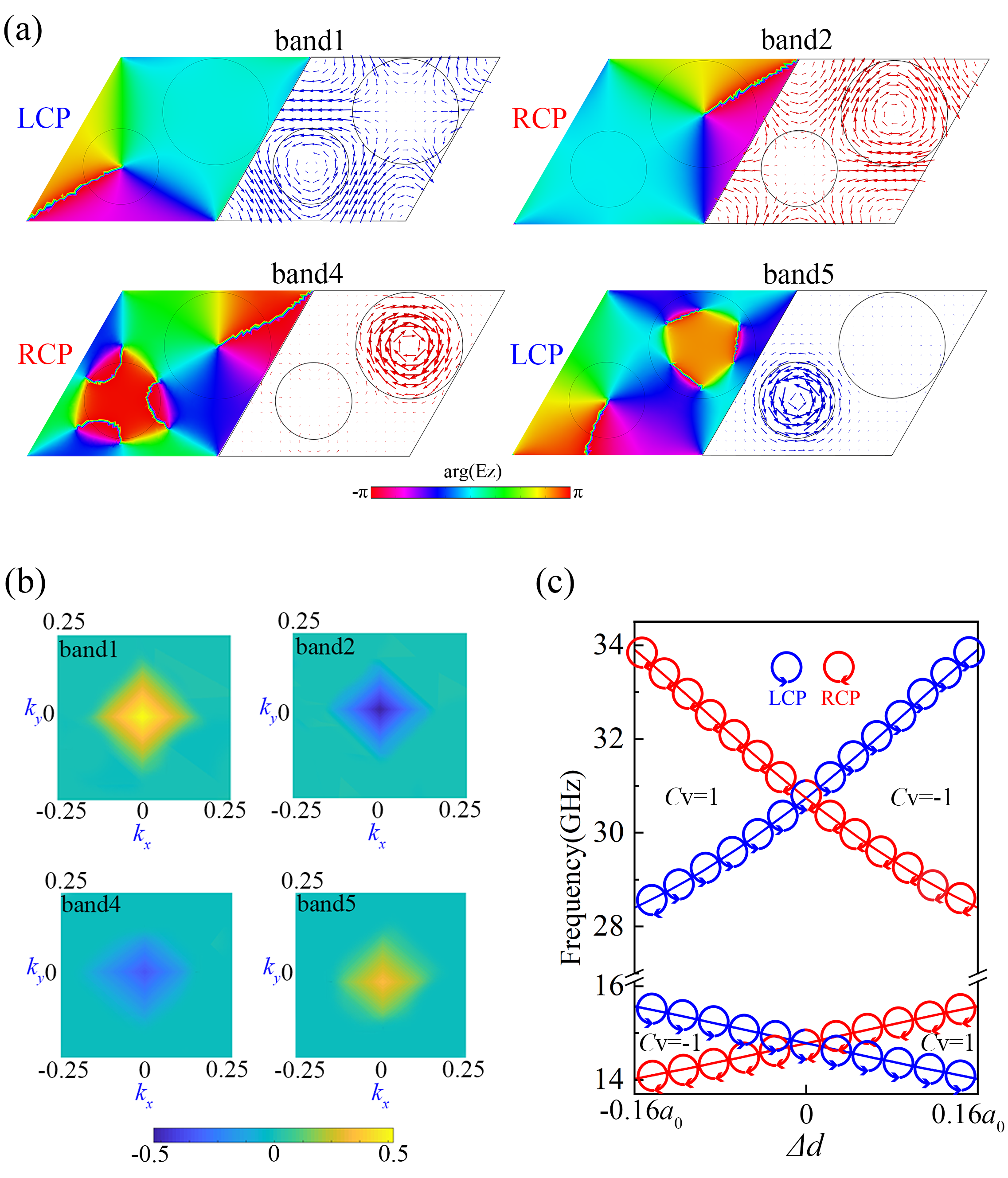} 
\caption{Field distributions and Berry curvatures of the valley-Hall PhC when $\Delta d=0.16a_0$. (a) Phases of $E_z$ and Poynting vector distributions of the four bands at the $K$ valley. For band 1 and band 5, the phases and Poynting vectors (blue arrows) indicate that the state is RCP. For band 2 and band 4, the state is LCP. (b) The local Berry curvature near the $K$ valley in momentum space. (c) The evolution of chirality at the $K$ valley by changing $\Delta d$. The phase is reversed at $\Delta d=0$ which indicates the phase transition point. The blue (red) arrows indicate LCP (RCP) states. The corresponding valley-Chern numbers ($C_{\mathrm{V}}$) are also reversed at $\Delta d=0$.}
\label{fig1}
\end{figure}
 
\subsection{\label{sec:level1}Emergence of hybrid wave-guiding modes}

\begin{figure*}[!htbp]
\centering\includegraphics[width=6.2in]{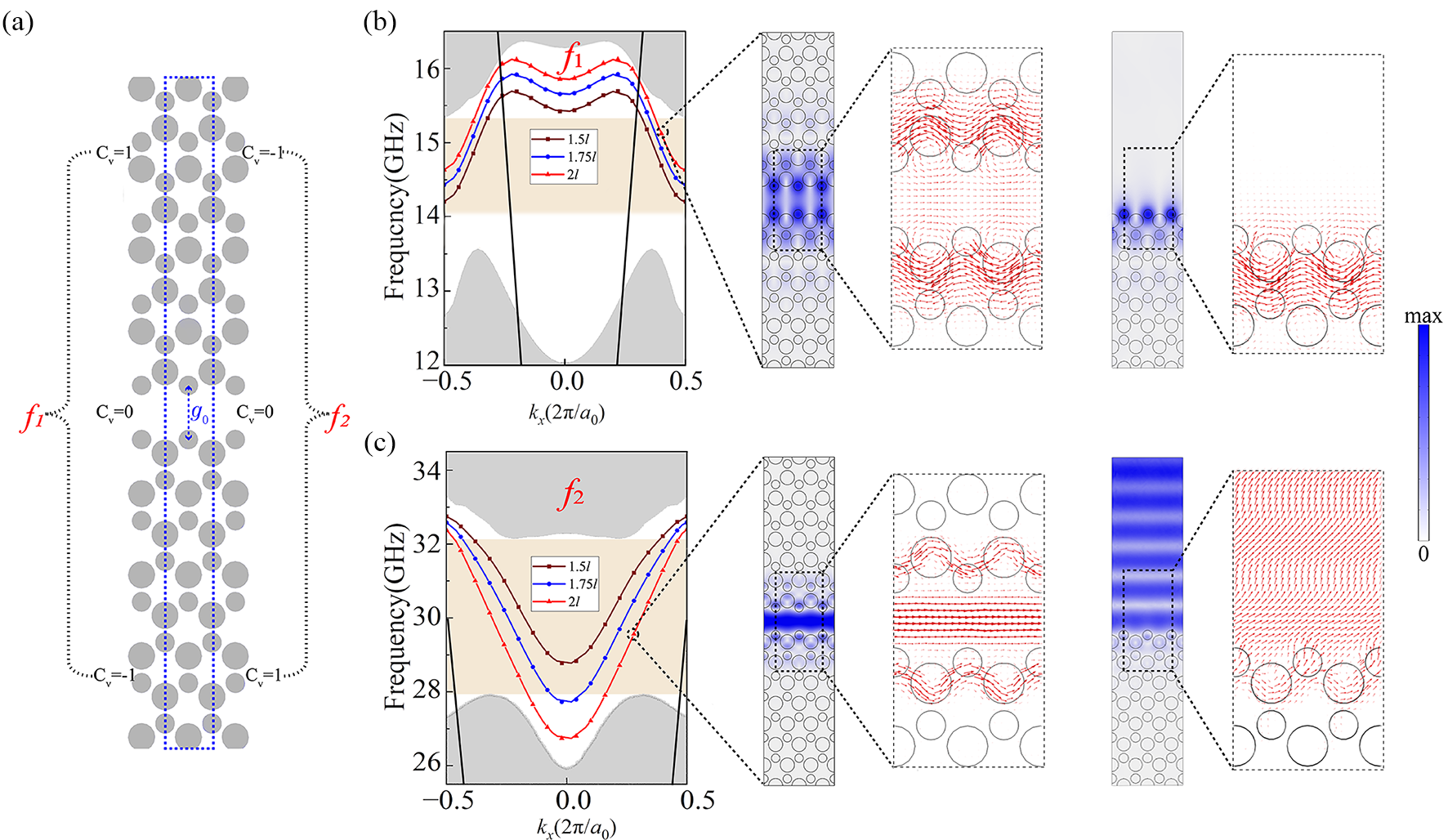} 
\caption{The proposed TVLW. (a) Supercell of the TVLW (blue dotted line). $g_0$ denotes the width of the air channel. For the valley-Hall PhCs, $\Delta d=\pm 1.6a_0$. (b, c) Left panels: the dispersions of wave-guiding states in $f_{\mathrm{1}}$ and $f_{\mathrm{2}}$ for different $g_0$. The orange areas indicate the frequency range of band gaps. The black lines indicate the light lines. Middle panels: distributions of electric fields and Poynting vectors of the modes at the marked locations when $g_0=2 l$. Right panels: distributions of electric fields and Poynting vectors of the modes when they are exposed to air.}
\end{figure*}

The nonzero valley-Chern number implies that the valley-Hall PhC can support nontrivial topological edge states. According to the bulk-edge correspondence\cite{2016SBulkSilveirinha}, topological edge states exist at the interface of two systems with different Chern numbers. Different from those methods by building the interfaces, here, we construct an air channel (can be considered to have a $C_{\mathrm{V}}=0$) with the width of $g_0$ sandwiched between two valley-Hall PhCs with opposite valley-Chern numbers, as shown in Fig.~3(a). The PhC with $\Delta d=0.16a_0$ ($\Delta d=-0.16a_0$) is arranged at the top (bottom). Remarkably, the width of the air gap in the proposed system is a freely adjustable parameter that can be used to tune the dispersion of the topological states, which cannot be achieved in the fixed interfaces that are constructed based on two distinct topological systems. As shown in the dispersion diagrams of Figs.~3(b) and (c), for fixed $g_0$, there is one band within each band gap. As $g_0$ increases, these states gradually move closer to the bulk states. The essential difference of the states within the two band gaps is that the states near the $K / K^{\prime}$ valley are below light lines (as indicated by the black lines) in $f_{\mathrm{1}}$ but are above the light lines in $f_{\mathrm{2}}$. Consequently, within $f_{\mathrm{1}}$, the states become surface edge states concentrated at the PhC-air domain walls. As can be seen from the distributions of electric fields and Poynting vectors, when $g_0=2 l$, within $f_{\mathrm{1}}$, the states below the light line are analogue to the surface states that are confined tightly in the dielectric region\cite{2020Local}. When they are exposed to air, their features can still be well preserved, as shown in the right panel of Fig. 3(b). The directions of helicity which correlate with the directions of the energy flow are opposite at different valleys, implying a valley-locked unidirectional propagation. Within $f_{\mathrm{2}}$, all states exist above the light line [Fig.~3(c)], indicating that there is a leakage of mode into the air region. For $g_0=2 l$, it can be seen that the electric fields are distributed not only at the PhC-air domain walls but also in the air channel. The Poynting vectors clearly show that a wave-guiding mode is generated in the core air channel and accompanied by the valley-locked energy transport along the PhC-air domain walls. Different from the modes within $f_{\mathrm{1}}$, these hybrid modes become radiative when there is no bound [right panel of Fig. 3(c)].

The analysis shows that the conditions for the emergence of the hybrid wave-guiding mode are as follows: 1) a nontrivial photonic bandgap, which is necessary for the generation of valley-locked edge states, and 2) the effective coupling between the topological edge states and air modes (i.e. the mode dispersion should be located above the light lines). With the above conditions fulfilled, the wave-guiding modes at the core air channel and the valley-locked edge states at the PhC-air domain walls are excited simultaneously. It is worth noting that the topology of the photonic bandgap is determined by the lattice of the structure and is independent of its local material and geometry parameters. The topological features of the hybrid modes will be explored in Section IV.

\section{\label{sec:level1}Interconnection with substrate integrated waveguides }
 
The proposed TVLW has an air channel with adjustable width, providing an additional degree of freedom for manipulating its eigenstates and making it much easier to interconnect with existing circuit components. Considering the similarity of the wave-guiding mode supported at the center air region of the TVLW with the conventional TE mode, we connect the TVLW with standard SIWs which support $\mathrm{TE}_{10}$ mode~\cite{2003SingleDeslandes,2005GuidedXu}, as shown in Fig.~4(a). The characteristics of the mode within the intersection area are shown in Fig.~4(b). The distribution of the magnetic-field lines illustrates that the mode within the air channel is a conventional $\mathrm{TE}_{10}$ mode. Therefore, the energy in the SIW can transit smoothly into the TVLW. Meanwhile, the distributions of Poynting vectors indicate that the dielectric rods at the edges of the TVLW act as scatterers that enable the effective mode conversion from the TE mode in the SIW to the hybrid mode in the TVLW. Figure 4(c) depicts the extracted propagation constant $\beta$ within the whole system at $29.4$~GHz. Evidently, a perfect momentum matching is obtained between SIWs and the TVLW. The coupling efficiency is evaluated by the reflection coefficient $S_{11}$ in Fig.~4(d). A reflection coefficient of lower than -10 dB is achieved within the whole operation frequency band ($f_{\mathrm{2}}$). These results illustrate that with the finite-width air channel, a highly efficient transition can be realized without designing complicated transit networks. Furthermore, conventional transmission lines (such as microstrip and coaxial lines) whose eigenmodes can couple to the $\mathrm{TE}_{10}$  mode can also be used to excite the hybrid mode effectively.

\begin{figure}[!htbp]
\centering\includegraphics[width=3.6in]{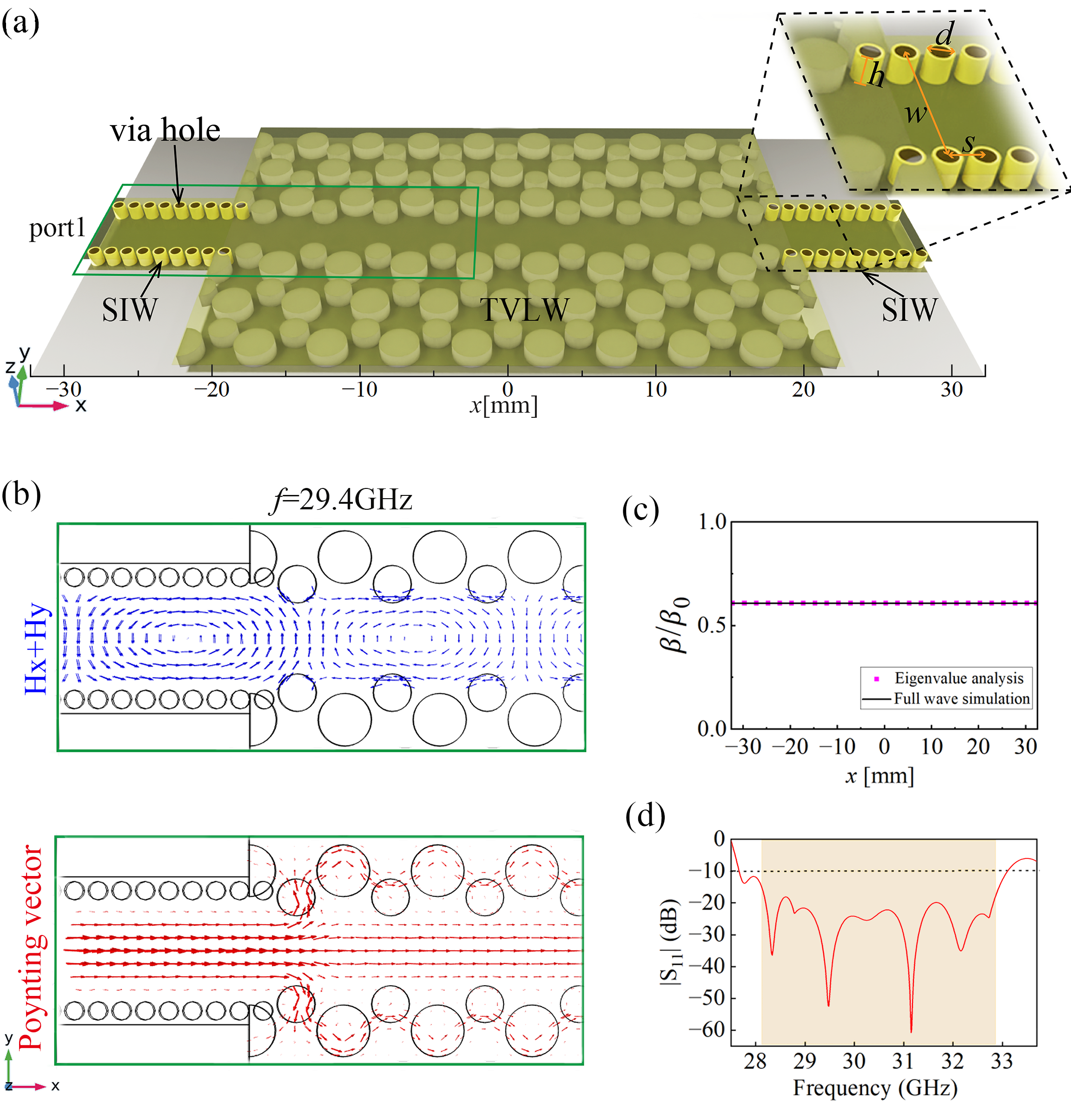} 
\caption{The TVLW interconnected with standard SIWs. (a) A schematic view. Parameters of the SIW: $h=1.08$~mm, $w=6.5$~mm, $d=1$~mm, and $s=1.25$~mm. (b) Distributions of magnetic-field lines and Poynting vectors at the intersection area from SIW to TVLW (the green area enclosed in (a)) when $f=29.4$~GHz. (c) The normalized propagation constant $\beta/\beta_0$ within the whole system at $29.4$~GHz. Here $\beta_0$ is the free-space propagation constant. (d) The simulated reflection coefficient, $S_{11}$ (the orange area indicates the frequency range of the second band gap).}
\end{figure}  

\begin{figure*}[!htbp]
\centering\includegraphics[width=5.5in]{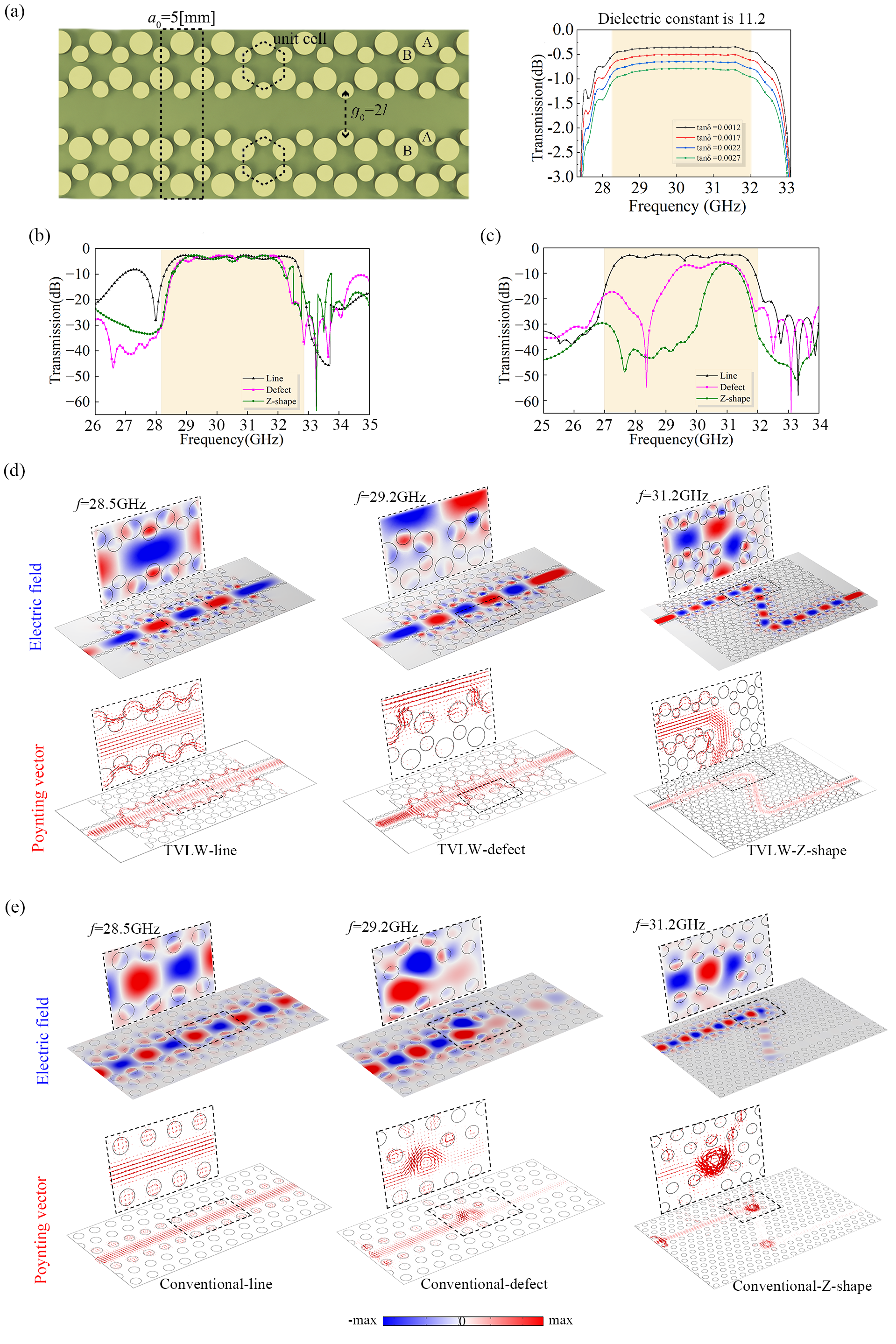} 
\caption{Robustness analysis of the proposed TVLW. (a) Its structural details and the transmission coefficients for the TVLW with a length of $45$~mm when different dielectric losses are introduced. Transmission coefficients of three different (straight, defective, and Z-shape) (b) TVLWs and (c) conventional PhC waveguides. (d) Distributions of electric fields and Poynting vectors of the corresponding three topological waveguides at different frequencies. (e) Distributions of electric fields and Poynting vectors of the corresponding three conventional PhC waveguides. The conventional waveguide is composed of dielectric cylinders of $r=1.4$~mm and a gap of $g_0=3l$. The working frequency is designed to be the same as the TVLW ($27-32$~GHz).}
\end{figure*}

\section{\label{sec:level1}Robustness analysis}

To verify the robustness of the proposed TVLW, we will conduct full-wave simulations for the higher band gap ($28-32$ GHz). When $g_0=2l$, the frequencies of the hybrid wave-guiding modes will occupy the entire band gap [red curve in Fig.~3(c)]. We first consider the influence of material loss. As shown in Fig. 5(a), when the loss factor ($\tan\delta$) increases by $0.0005$, for the TVLW with a length of $45$ mm, the transmission coefficient will decrease by approximately $0.14$ dB.

Then, we explore the performance of the TVLW by introducing structural defects. Three different (straight, defective, and Z-shape) waveguides are constructed based on the proposed TVLW and conventional PhC waveguide, respectively. The simulated transmission coefficients of the total six waveguides are shown in Figs.~5(b) and (c). The transmission loss is less than $3$~dB within the whole operation frequency band for each of the TVLWs. On the contrary, the defective and Z-shaped conventional PhC waveguides have transmission losses of more than $10$~dB. They encounter strong scattering and reflection when there are defects and bends [Fig. 5(e)]. Figure 5(d) shows the electric-field distributions and Poynting vectors of the straight line, defective, and Z-shaped topological waveguides at different frequencies. Obviously, the hybrid wave-guiding modes supported in the TVLWs present the features of topological protection with robust EM transmission against defects and bends. In the defective waveguide, the valley-locked edge states can bypass the defect (black dotted rectangle) and the hybrid mode is reconstructed without backscattering. Meanwhile, when the EM waves propagate along the Z-shaped bend, there is no loss, and the hybrid mode can smoothly bypass the 60-degree bend. 
Therefore, it is confirmed that the wave-guiding modes supported in the designed TVLWs possess the properties of topological protection as in valley-Hall PhCs. In the presence of the conventional TE modes in air channels, the concept of topological robustness has been successfully incorporated into conventional waveguiding modes, which provides an innovative platform for integrating topological protection effects into existing circuits.

\section{Conclusion}

In summary, a new TVLW by decorating the surfaces of a conventional waveguide with valley-Hall PhCs has been proposed. By enabling the coupling of air modes and topological edge states, a hybrid mode consisting of a TE mode and valley-locked edge states has been generated. Most importantly, the feature of topological protection of the hybrid modes has been demonstrated in mmWave regime. Benefiting from the adjustable width of the proposed TVLW, an efficient intersection with standard photonic circuit components has been achieved. Our wave-guiding mechanism can be used in general cases such as feeding/matching network design, high-capacity energy transport, and high-quality signal transmission in mmWave integrated circuits.

\section*{Acknowledgments}
R. Z. and M. L. N. C. are supported by the National Natural Science Foundation of China under Grant 62301470 and the Hong Kong Polytechnic University Start-up fund for RAPs under Grant BD2P. Y. L. and Y. T. thank Prof. Zhou Bin, and are supported by the Young Scientist Fund [NSFC11804087], National Natural Science Foundation of China [NSFC12247101], and Science and Technology Department of Hubei Province [2022CFB553]. R. Z., X. S., Y. R., Z. Y., and H. L. are supported by the Key Research and Development Proram of Hubei Province under Grant [2023BAB061] and in part by the Fundamental Research Funds for the Central Universities.


\begin{thebibliography}{00}

\bibitem{2021ShenmmW} G. Shen, W. Feng, W. Che, Y. Shi, and Y. Shen, “Millimeter-wave dual-band bandpass filter with large bandwidth ratio using GaAs-based integrated passive device technology,” \emph{IEEE Electron Device Lett.}, vol. 42, no. 4, pp. 493-496, Apr. 2021,

\bibitem{2022GaOn}Z. Ge, L. Chen, L. Yang, R. Gómez-Garcia, and X. Zhu, “On-chip millimeter-wave integrated absorptive bandstop filter in (Bi)-CMOS technology,” \emph{IEEE Electron Device Lett.}, vol. 42, no. 1, pp. 114-117, Jan. 2020.

\bibitem{2019BautistaCompact} M. G. Bautista, H. Zhu, X. Zhu, Y. Yang, Y. Sun, and E. Dutkiewicz, “Compact millimeter-wave bandpass filters using quasi-lumped elements in 0.13-$\mu$m (Bi)-CMOS technology for 5G wireless systems,” \emph{IEEE Trans. Microw. Theory Techn.}, vol. 67, no. 7, pp. 3064-3073, Jul. 2019.

\bibitem{2003SingleDeslandes} D. Deslandes, and K. Wu,  
 “Single-substrate integration technique of planar circuits and waveguide filters,” \emph{IEEE Trans. Microw. Theory Tech.}, vol. 51, no. 2, pp. 593-596, Feb. 2003.

\bibitem{2005GuidedXu} F. Xu and W. Ke, “Guided-wave and leakage characteristics of substrate integrated waveguide,”
\emph{IEEE Trans. Microw. Theory Tech.}, vol. 53, no. 1, pp. 66-73, Jan. 2005.


\bibitem{2015DesignBrazalaz} A. A. Brazalez, E. Rajo-Iglesias, J. L. Vazquez-Roy, A. Vosoogh, and P. S. Kildal, “Design and validation of microstrip gap waveguides and their transitions to rectangular waveguide, for Millimeter-Wave Applications,” \emph{IEEE Trans. Microw. Theory Tech.}, vol. 63, no. 12, pp. 4035-4050, Dec. 2015.

\bibitem{2014BroadbandMa} H. F. Ma, X. P. Sheng, Q. Cheng,  W. X. Jiang, and T. J. Cui, “Broadband and high-efficiency conversion from guided waves to spoof surface plasmon polaritons,” \emph{Laser Photonics Rev.}, vol. 8, no. 1, pp. 146-151, Jan. 2014.

\bibitem{2019AdiabaticXu} Z. X. Xu, X. Y. Yin, and  D. F. Sievenpiper, “Adiabatic mode-matching techniques for coupling between conventional microwave transmission lines and one-dimensional impedance-interface waveguides,” \emph{Phys. Rev. Appl.}, vol. 11, no. 004071, pp. 1-6, Apr. 2019.

\bibitem{2017StudyZhang} J. Zhang, X. Zhang, and A. A. Kishk, “Study of bend discontinuities in substrate integrated gap waveguide,” \emph{IEEE Microw. Wirel Compon. Lett.}, vol. 27, no. 3, pp. 221-223, Mar. 2017.

\bibitem{2013PhotonicKhanikaev} A. B. Khanikaev, S. H. Mousavi, W. K. Tse, M. Kargarian, A. H. MacDonald, and G. Shvets, “Photonic topological insulators,” \emph{Nat. Mater.}, vol. 12, no. 3, pp. 233-239, Mar. 2013.

\bibitem{2008PossibleHaldane} F. D. M. Haldane and S. Raghu, “Possible realization of directional optical waveguides in photonic crystals with broken time-Reversal symmetry,” \emph{Phys. Rev. Lett.}, vol. 100, no. 1, pp. 013904, Jan. 2008.
  
\bibitem{2008ReflectionWang} Z. Wang, Y. D. Chong, John D. Joannopoulos, and M. Solja\ifmmode \check{c}\else \v{c}\fi{}i\ifmmode \acute{c}\else \'{c}\fi{}, “Reflection-free one-way edge modes in a gyromagnetic photonic crystal,” \emph{Phys. Rev. Lett.}, vol. 100, no. 1, pp. 013905, Jan. 2008.
  
   
\bibitem{2015SchemeWu} L. H. Wu and X. Hu, 
 “Scheme for achieving a topological photonic crystal by using dielectric material,” \emph{Phys. Rev. Lett.}, vol. 114, no. 22, pp. 223901, Jun. 2015.

\bibitem{2016ObservationLu} J. Y. Lu, C. Y. Qiu, L. P. Ye, X. Y. Fan, M. Z. Ke, F. Zhang, and Z. Y. Liu, “Observation of topological valley transport of sound in sonic crystals,” \emph{Nat. Phys.}, vol. 13, no.4, pp. 369-374, Apr. 2017.

\bibitem{2022HigherZhou} R. Zhou, H. Lin, Y. J. Wu, Z. F. Li, Z. H. Yu, Y. Liu, and D. H. Xu, “Higher-order valley vortices enabled by synchronized rotation in a photonic crystal,” \emph{Photonics Res.}, vol. 10, no. 5, pp. 1244-1255, May. 2022.
   
\bibitem{2022lowLi} Z. F. Li, H. Lin, R. Zhou, X. T. Shi, Z. H. Yu, Y. Liu, Y. J. Wu, and R. X. Tang, “A low-loss zero-index photonic crystal slab based on toroidal dipole mode,” \emph{Opt. Express}, vol. 30, no. 14 pp. 25544--25554, Jul. 2022.

\bibitem{2011ExperimentalPoo}Y. Poo, R.-X. Wu, Z. F. Lin, Y. Yang, and C. T. Chan, “Experimental realization of self-guiding unidirectional electromagnetic edge states,” \emph{Phys. Rev. Lett.}, vol. 106, no. 9, pp. 093903, Mar. 2011. 

\bibitem{2017ValleyleChen} X. D. Chen, F. L. Shi, M. Chen, and J. W. Dong, “Valley-contrasting physics in all-dielectric photonic crystals: Orbital angular momentum and topological propagation,” \emph{Phys. Rev. B}, vol. 96, no. 2, pp. 020202, Jul. 2017.

\bibitem{2018TunableChen} X. D. Chen, F. L. Shi, H. Liu, J. C. Lu, and J. W. Dong, “Tunable electromagnetic flow Control in valley photonic crystal waveguides,” \emph{Phys. Rev. Appl.}, vol. 10, no. 4, pp. 1-8, Oct. 2018.

\bibitem{2018YangVisualization} Y. T. Yang, Y. F. Xu, T. Xu, H-X Wang, J-H Jiang, X. Hu, and Z. H. Hang, “Visualization of a unidirectional electromagnetic waveguide using topological photonic crystals made of dielectric materials,” \emph{ Phys. Rev. Lett.}, vol. 120, no. 21, pp. 1-6, May. 2018.

\bibitem{2021TopologicalZhou} R. Zhou, H. Lin, H. Y. Liu, X. T. Shi, R. X. Tang, Y. J. Wu, and Z. H. Yu, “Topological edge states of Kekulé-type photonic crystals induced by a synchronized rotation of unit cells,” \emph{Phys. Rev. A}, vol. 104, no. 3, pp. L031502, Sep. 2021.

\bibitem{2019PseudospinChen} M. L. Chen, L. J. Jiang, Z. H. Lan, and  W. Sha, “Pseudospin-polarized topological line defects in dielectric photonic crystals,” \emph{IEEE Trans. Antennas Propag.}, vol. 68, no. 1 pp. 609-613, Jan. 2020.
   
 
\bibitem{2022TopologicalYu} Z. H. Yu, H. Lin, R. Zhou, Z. F. Li, Y. Mao, K. Peng, Y. Liu, and X. T. Shi, “Topological valley crystals in a photonic Su--Schrieffer--Heeger (SSH) variant,” \emph{J. Appl. Phys.}, vol. 132, no. 16 pp. 163101, Oct. 2022.

\bibitem{2020TerahertzYang} Y. H. Yang, Y. Yamagami, X. B. Yu, P. Pitchappa, J. Webber, B. L. Zhang, M. Fujita, T. Nagatsuma, and R. Singh, “Terahertz topological photonics for on-chip communication,” \emph{Nat. Photonics}, vol. 14, no. 7 pp. 446-451, Jul. 2020.

\bibitem{2019TopologicalMa} J. W. Ma, X. Xi, and X. K. Sun, “Topological photonic integrated circuits based on valley kink states,” \emph{Laser Photonics Rev.}, vol. 13, no. 12, pp.1-7, Dec. 2019.

\bibitem{2017ValleyDong} J. W. Dong, X. D. Chen,  H. Zhu, Y.  Wang, and X. Zhang, “Valley photonic crystals for control of spin and topology,” \emph{Nat. Mater.}, vol. 16, no. 3, pp. 298-302, Mar. 2017.
      
\bibitem{2018ValleyChenQ} Q. L. Chen, L. Zhang, M. J. He, Z. J. Wang, X. Lin, F. Gao, Y. H. Yang, B. Zhang, and H. S. Chen, “Valley-Hall photonic topological insulators with dual-band kink states,” \emph{Adv. Opt. Mater.}, vol. 7, no. 15, pp. 1-5, Agu. 2019.

\bibitem{2011DirectWu} X. X. Wu, Y. Mang, J. X. Tian, Y. Z. Huang, H. Xiang, D. Z. Han, and W. J. Wen, “Direct observation of valley-polarized topological edge states in designer surface plasmon crystals,” \emph{Nat. Commun.}, vol. 8, no. 1, pp. 1304, Nov. 2017.
    
\bibitem{2018ValleyZhang} L. Zhang, Y. H. Yang, M. J. He, H. X. Wang, Z. J. Yang, E. P. Li, F. Gao, B. Zhang, J. H. Jiang, and H. S. Chen, “Valley kink states and topological channel intersections in substrate-integrated photonic circuitry,” \emph{Laser Photonics Rev.}, vol. 13, no. 11, pp. 1-7, Nov. 2019.
 
\bibitem{2022ValleyLi} B. L. Li, H. Y. Shi, W. Sha, J. J. Yi, G. Q. Li, A. X. Zhang, and Z. Xu, “Valley topological line-defects for Terahertz waveguides and power divider,” \emph{Opt. Mater.}, vol. 126, no. 112152 pp. 1-8, Apr. 2022.


\bibitem{2020ValleyWang} M. D. Wang, W. Y. Zhou, L. Y. Bi, C. Y. Qiu, M. Z. Ke, and Z. Y. Liu, “Valley-locked waveguide transport in acoustic heterostructures,” \emph{Nat. Commun.}, vol. 11, no. 1, pp. 1-6, Jun. 2020.

\bibitem{2022ExtendedWang} J. Q. Wang, Z. D. Zhang, S. Y. Yu, H. Ge, K. F. Liu, T. Wu, X. C. Sun, L. Liu, H. Y. Chen, C. He, M. H. Lu, and Y. F. Chen, “Extended topological valley-locked surface acoustic waves,” \emph{Nat. Commun.}, vol. 13, no. 1, pp. 1-8, Mar. 2022.

 
\bibitem{2021TopologicalWang} M. D. Wang, R. Y. Zhang, L. Zhang, D. Y. Wang, Q. H. Guo, Z. Q. Zhang, and C. T. Chan, “Topological one-way large-area waveguide states in magnetic photonic crystals,” \emph{Phys. Rev. Lett.}, vol. 126, no. 6, pp. 1-6, Feb. 2021.


\bibitem{2021PhotonicChen} Q. L. Chen, L. Zhang, F. J. Chen, Q. H. Yan, R. Xi, H.  S. Chen, and Y. H. Yang, “Photonic topological valley-locked waveguides,” \emph{ACS Photonics}, vol. 98, no. 6, pp. 1-10, Jun. 2021.


\bibitem{2023ExtendedZhao} Y. L. Zhao, F. Liang, J. F. Han, D. H. Zhao, and B.-Z. Wang, “Extended terahertz valley-locked surface waves in designer surface plasmon crystals,” \emph{J. Phys. D: Appl. Phys.}, vol. 56, no. 18, pp. 1-10, Mar. 2023.


\bibitem{2023Coexistenceli} S. J. Li, M. L. Chen, Z. H. Lan, and P. Li, “Coexistence of large-area topological pseudospin and valley states in a triband heterostructure system,” \emph{Opt. Lett.}, vol. 48, no. 17, pp. 1-5, Sept. 2023.

  
\bibitem{2019QuantizationBenalcazar} W. A. Benalcazar, T. Li, and T. L. Hughes, “Quantization of fractional corner charge in Cn-symmetric higher-order topological crystalline insulators,” \emph{Phys. Rev. B}, vol. 99, no. 24, pp. 3-20, Jun. 2019.

\bibitem{2011DirectOkada} Y. Okada, C. Dhital, W. Zhou, E. D. Huemiller, H. Lin, S. Basak, A. Bansil, Y. B. Huang, H. Ding, and  Z. Wang, “Direct observation of broken time-Reversal symmetry on the surface of a magnetically doped topological insulator,” \emph{Phys. Rev. Lett.}, vol. 106, no. 20, pp. 2-4, May. 2011.

\bibitem{2020FirstZhao} R. Zhao, G. D. Xie, M. L. Chen, Z. H. Lan,  and  W. Sha, “First-principle calculation of Chern number in gyrotropic photonic crystals,” \emph{Opt. Experss}, vol. 28, no. 4, pp. 4638-4649, Feb. 2020.

\bibitem{2016SBulkSilveirinha} M. G. Silveirinha, “Bulk-edge correspondence for topological photonic continua,” \emph{Phys. Rev. B}, vol. 94, no. 20, pp. 1-11, Nov. 2016.

\bibitem{2020Local} M. L. Chen, L. J. Jiang, Z. H. Lan, and W. Sha, “Local orbital-angular-momentum dependent surface states with topological protection,” \emph{Opt. Experss}, vol. 28, no. 10, pp. 14428-14435, May. 2020.
   
 

\end{thebibliography}
\end{document}